\documentclass{ws-procs975x65}            
\usepackage{epstopdf}
\begin{document}
\title{Measurement of the Muon content of EAS with the Pierre Auger Observatory}

\author{J. C. Espadanal$^{1}$ for the Pierre Auger Collaboration$^{2}$}

\address{$^{1}$LIP - Laborat\'orio de Instrumenta\c{c}\~{a}o e F\'isica Experimental de Part\'iculas,\\
Lisbon, 1000-149, Portugal \\
$^*$E-mail: jespada@lip.pt}

\address{$^{2}$Observatorio Pierre Auger, Av. San Mart\'in Norte 304, 5613 Malarg\"ue, Argentina\\
(Full author list: http://www.auger.org/archive/authors\_2014\_06.html) }

\begin{abstract}
Several methods developed within the Pierre Auger Collaboration for the estimation of the muonic component of the Extensive Air Showers observed in the surface Cherenkov detectors are described. The results derived from the data show a deficit of muons predicted by the current hadronic interactions models at ultra-high energies.
\end{abstract}


\bodymatter

\section{Introduction}
The Pierre Auger Observatory is a hybrid detector with the aim of studying the characteristics of the longitudinal and lateral profiles of extensive air showers (EAS) from ultra high energy cosmic rays (UHECR) with a fluorescence detector array (FD) and a surface detector array (SD), respectively \cite{PAO,FArqueiros}. One of the goals is to determine the mass composition of the UHECR. 
An important feature to distinguish composition is the muon content of the shower since heavier primaries will generate more muons for the same energy. The predicted number of muons depends on the hadronic model used.
The water-Cherenkov detectors measure only a combination of the much larger electromagnetic component and the muonic component reaching the ground, but different techniques are used to recover the muonic content.

We present the results of three different classes of methods: 1) separating the muonic and electromagnetic signals based on their different time structures, 2) using inclined showers, for which the electromagnetic component has been strongly attenuated in crossing the atmosphere and 3) comparing the ground signals with the longitudinal profile, for hybrid events seen also with the FD. Data are compared with the hadronic interaction models QGSJETII-04\cite{QGS} and EPOS-LHC\cite{EPOS}.

\section{Estimate of the Muonic Signal in Data}
The Cherenkov photons produced by the shower particles in the SD water-Cherenkov detectors (WCD) are sampled by three photomultipliers and digitized with FADCs in $25$ ns bins. The time response of individual particles is the same for muonic and electromagnetic (EM) particles. Nevertheless, there are two features that enable us to separate each component: the signal amplitude and the time-of-arrival distribution. The energy deposition in WCD for muons ($\approx240$ MeV) is much higher than for electrons and photons ($\approx10$ MeV). The number of EM particles is, on average, one order of magnitude higher than muons (at relatively low zenith angles). The muon signals are peaked and short and the EM signals are smooth and elongated. Concerning the time-of-arrival, the muons arrive earlier than the EM particles. 
Sections \ref{smooth} and \ref{multivariate} exploit those features. These methods have some limitations that gives rise to variances and systematic bias between models and primaries (due to muon pile-up, small muon peaks due to corner clipping on the WCD, signal fluctuations). The main source of uncertainty comes from the high energy photons that can produce a signal similar to muons. Their contribution is expected to be less than $10\%$ to $15\%$ for proton- and iron-induced showers in the energy and angular ranges of the methods.

\subsection{Smoothing method}
\label{smooth}
The EM signal ($S_{\text{EM}}$) is much smoother than the muon signal, so smoothing the total signal by averaging over $N_{bins}$ bins, provides an estimation for the electromagnetic contribution. The difference to the original trace is assigned to be the muonic component $S_{\mu}$. The procedure is repeated for a number of iterations $N_{iter}$. To avoid under- or over-smoothing, the smoothing window, $N_{bin}$, and $N_{iter}$ are optimized with Monte Carlo simulations to minimize the bias and variances. The overall bias obtained for muon fractions for different models and primaries is about $\pm0.05$ and the average resolution is $\pm0.08$ \cite{ICRC13-Kegl}.

\subsection{Multivariate method}
\label{multivariate}
The basic idea of the method is to combine the characteristics of the FADC signal in a equation to estimate the muon fraction $f_\mu$. The function was chosen to be
\begin{equation}
\hat{f}_{\mu}=a+b\hat{\theta} +cf^{2}_{0.5} + d \hat{\theta} P_{0} + e \hat{r}
\label{eq: multi}
\end{equation}
where $\hat{\theta}$ is the reconstructed zenith angle of the shower, $\hat{r}$ is the distance of the WCD to the shower axis, $f_{0.5}$ is the portion of the signal larger than 0.5 VEM\footnote{Vertical Equivalent Muon} given by $f_{0.5}=\frac{1}{S} \sum^{N}_{j=1} x_j II\{x_j>0.5\}$, $II\{A\}$ is 1 if A is true and 0 otherwise, $x_j$ is the signal of the $j$ bin in the signal vector trace $\textbf{x}=(x_1,...x_N)$. The $P_0$ function is the normalized zero-frequency component of the power spectrum given by,
\begin{equation}
P_0 = \frac{S^2}{N\sum^N_{j=1}x_j^2} = \frac{\langle \textbf{x} \rangle^2}{\langle \textbf{x}^2\rangle} = \left[ 1+ \frac{\sigma^2(\textbf{x})}{\langle\textbf{x}\rangle^2}\right]^{-1}
\label{eq: multi2}
\end{equation}
where $\langle \textbf{x} \rangle = S/N$ is the mean of the signal, $\sigma^2(\textbf{x})$ is the variance of the signal vector and $\langle \textbf{x}^2 \rangle$ is the second moment. The $f_{0.5}$ and $P_0$ are sensitive to large relative fluctuations and short signals. The parameters ($a,b,c,d,e$) were fitted to Monte Carlo simulations. There are other kinds of parametrizations similar to eq \ref{eq: multi}, but this one was chosen with the objective of minimizing the variance and sensitivity of the estimator $\hat{f}$ to composition and models. The overall bias in $\hat{f_\mu}-f_\mu$ for the different primaries and models is $\pm0.02$ and the average resolution is $\pm0.08$ \cite{ICRC13-Kegl}.

\subsection{Inclined showers}
\label{Inclined}
In very inclined showers, at zenith angles above $60^{\circ}$, the ground signal in the SD is mainly from muons, since the other components have been absorbed higher in the atmosphere. The ground signal can be fitted to the equation,
\begin{equation}
\rho = N_{19}  \rho_{\mu,19}(x,y,\theta,\phi)
\label{eq: inclined}
\end{equation}
$\rho_{\mu,19}$ is the model prediction\cite{Dembinski} normalized to the reference muon density at the ground for proton showers of $10^{19}$ eV with QGSJetII-03 and FLUKA interaction models. $N_{19}$ is the shower size parameter, that gives the number of muons relative to the reference and can be used as an energy estimator for the shower in this zenith angle range. We want the muon content, so we compare $N^{\text{MC}}_{19}=N_{\mu}/N_{\mu,19}$ with the fractions of the total number of muons on the ground $R^{\text{MC}}_{19}=N^{true}_{\mu}/N_{\mu,19}$, at Monte Carlo level ($N_{\mu,19}$ is the total number of muons for the reference). For several models and primaries and get a resolution of $8\%$ and a systematic uncertainty smaller than $5\%$. Finally we can look at the parameter $R_{\mu}$, which is $N_{19}$ corrected with this average bias to $R^{\text{MC}}_{19}$ (see \cite{ICRC13-Valino}).

\section{Application to data}
The three methods described above were applied to the data collected between 01 January 2004 and 31 December 2012. A deficit in data was found for the number of muons predicted from simulations assuming different hadronic interaction models.\\
The smoothing and multivariate methods were applied to events with zenith angles $\hat{\theta}<60^{\circ}$ and energy $\hat{E}\in\left[10^{18.98},10^{19.02}\right]$ eV. Only the detectors with a distance from the shower axis $\hat{r}\in\left[950,1050\right]$ m were used. At $\hat{E}=10^{19}$ eV the resolutions on the core position and energy are about $50$ m and $12\%$, and the energy systematic uncertainty is $14\%$. In figure \ref{fig: muons1}(a), we have the results as a function of zenith angle for these two methods. The average muon signal found in data divided by the muon signal in QGSJetII-04, for a proton shower at $10^{19}$ eV, at $1000$ m and $\theta<60^{\circ}$ is:\\
$1.33\pm0.02(stat.)\pm0.05(sys.) \quad (multivariate)$\\
$1.31\pm0.02(stat.)\pm0.09(sys.) \quad (smoothing)$\\

For the inclined shower analysis, the data were hybrid events with zenith angles $62^{\circ}<\hat{\theta}<80^{\circ}$ with an adapted version of the FD quality cuts for these zenith angles \cite{ICRC13-Valino}. The results of the parameter $R_\mu$ as function of energy are plotted in figure \ref{fig: muons1}(b). At $10^{19}$eV, the number of muons in the data exceeds the one from proton (iron) simulations with QGSJetII-03 by a factor of $1.8$ ($1.4$). Note that QGSJetII-04 and EPOS-LHC have $20\%$ and $30\%$ more muons, at this energy, than QGSJETII-03.

\def\figsubcap#1{\par\noindent\centering\footnotesize(#1)}
\begin{figure}[h]%
\begin{center}
  \parbox{2.1in}{\includegraphics[width=2.15in]{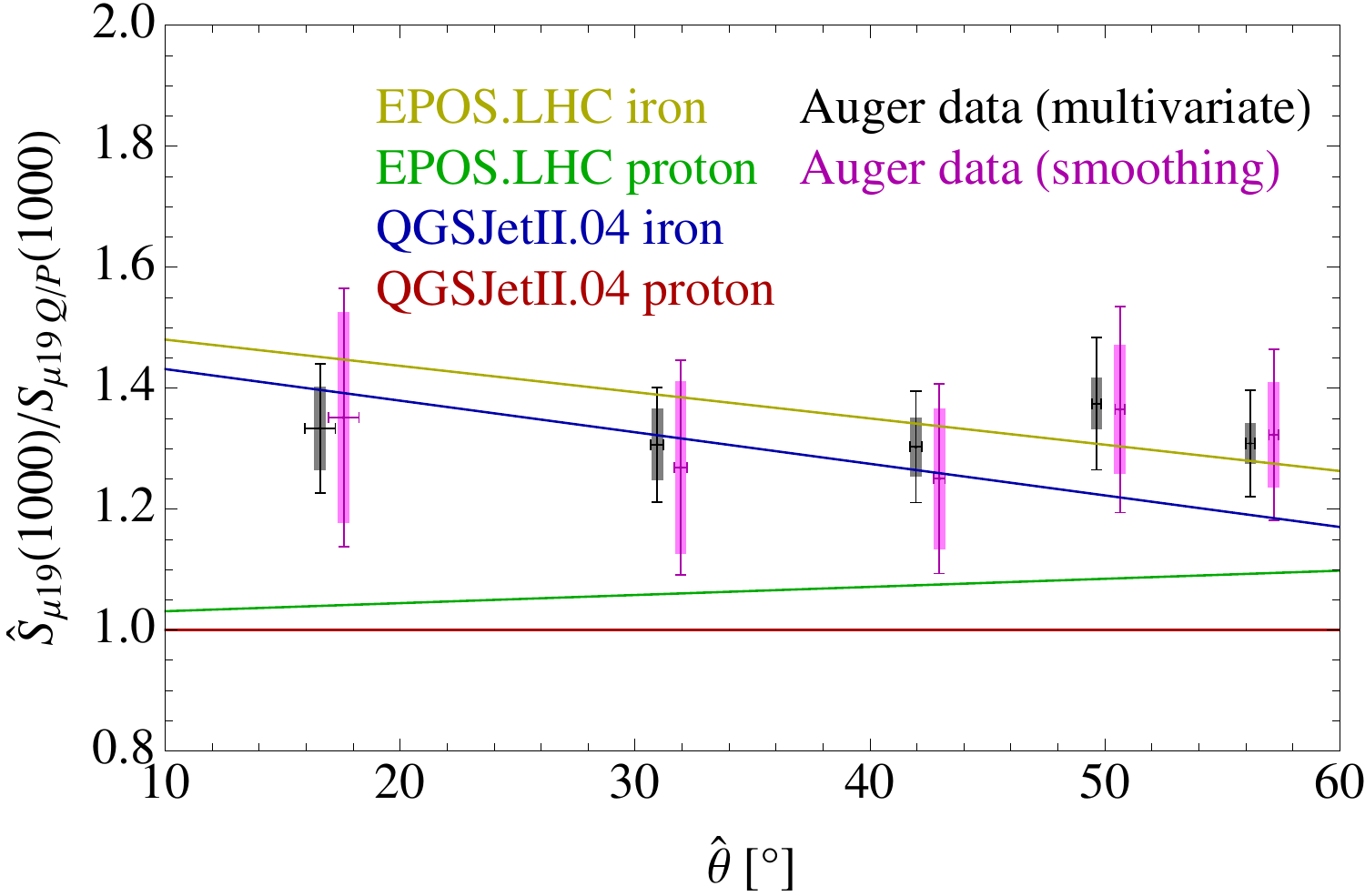}\figsubcap{a}}
  \hspace*{4pt}
  \parbox{2.1in}{\includegraphics[width=1.55in]{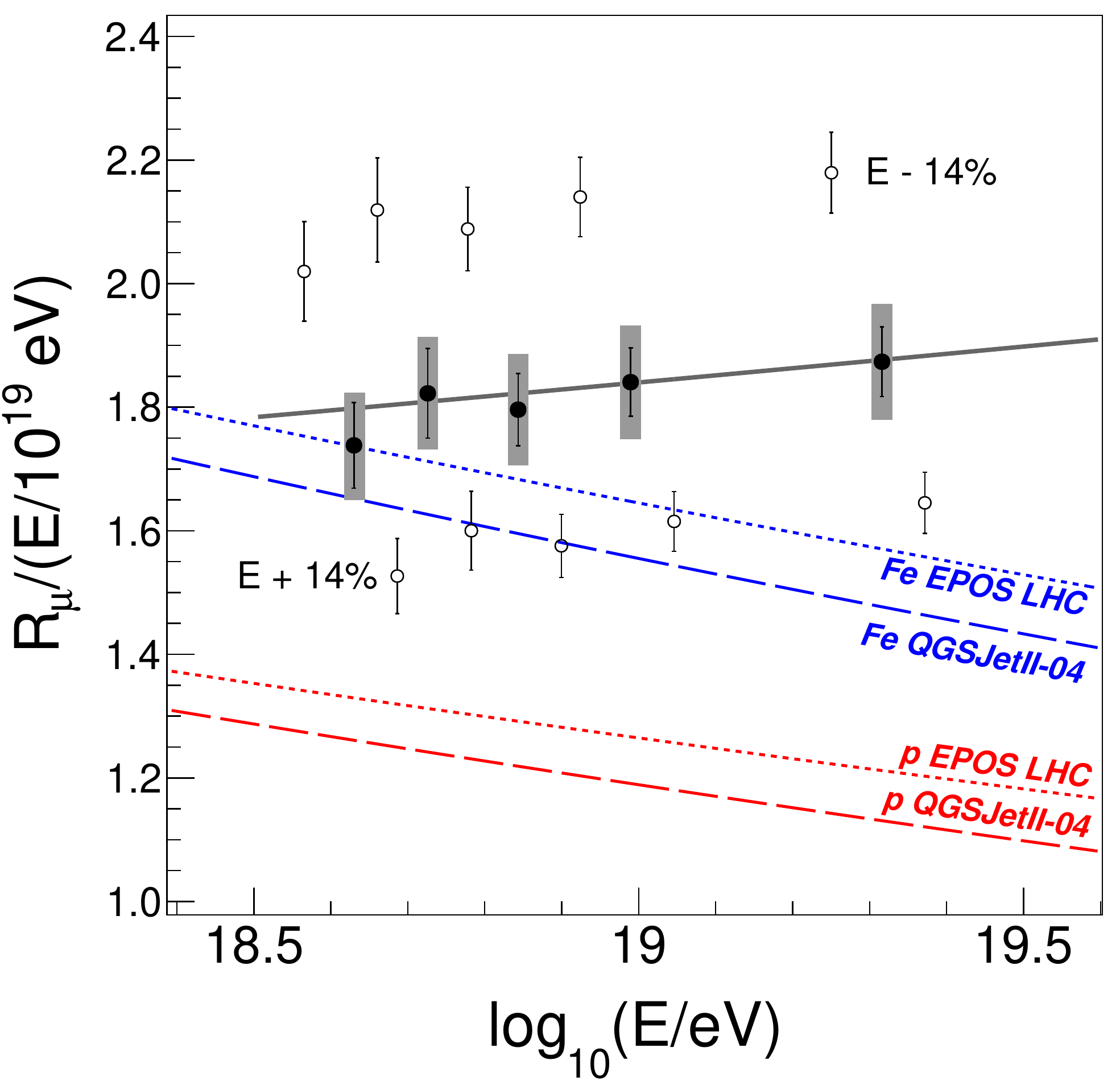}\figsubcap{b}}
  \caption{(a) Muon signal measured as a function of zenith angle for the smoothing and multivariate method. Values of the muon rescaling for an energy $10^{19}$ eV,  at $1000$ m from shower axis with respect to QGSJetII-04. The error bars and the rectangles are the statistical and systematic uncertainties. Points are displaced by $\pm0.5^{o}$ for visibility. (b) Average value of 
  $R_\mu /(E_{\text{FD}}/10^{19}\text{eV})$ as function of the FD energy, with relation to QGSJetII-03. The gray rectangles are the syst. uncertainties of $R_\mu$. The open circles correspond to the $R_\mu$ if $E_{\text{FD}}$ is varied by its syst. uncertainty.}%
  \label{fig: muons1}
\end{center}
\end{figure}

\section{Study of hybrid events}
\begin{figure}[h]%
\begin{center}
  \parbox{2.1in}{\includegraphics[width=1.75in]{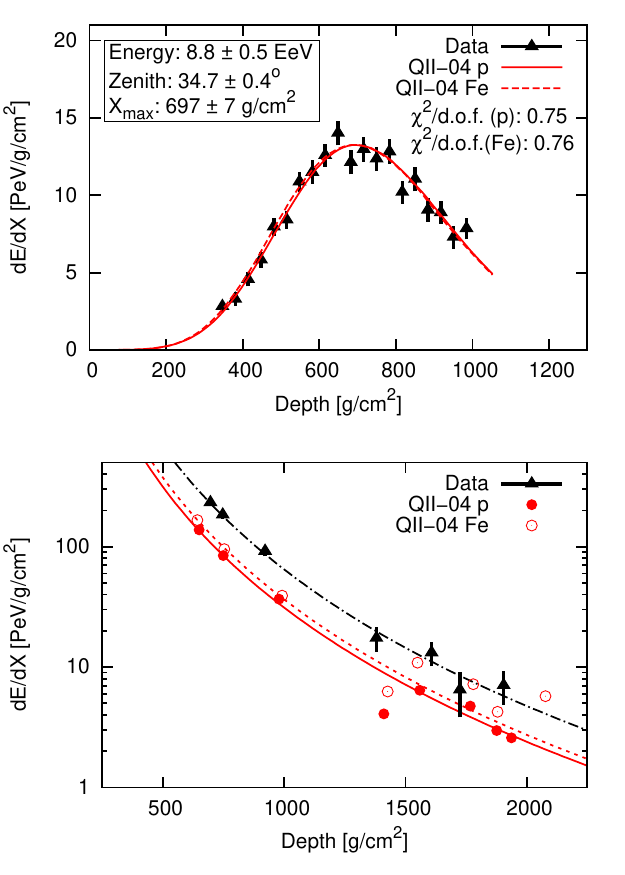}\figsubcap{a}}
  \hspace*{4pt}
  \parbox{2.1in}{\includegraphics[width=2in]{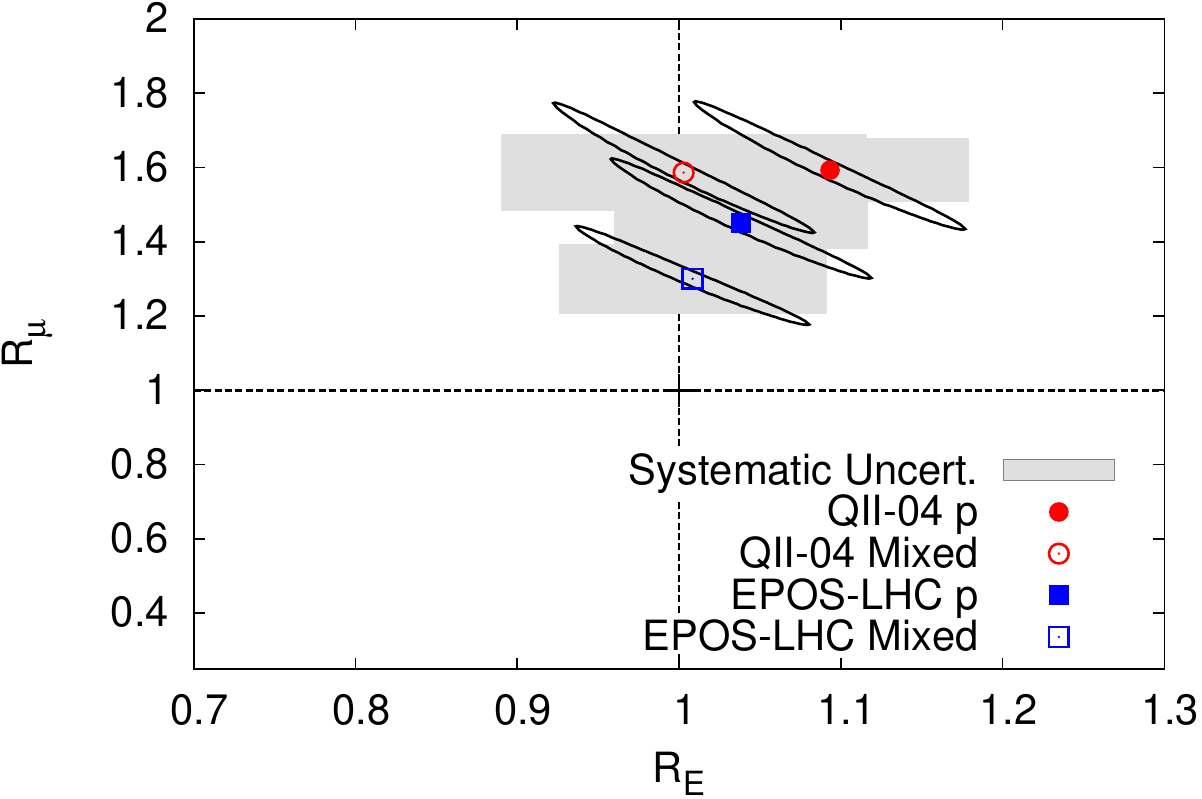}\figsubcap{b}}
  \caption{Left: (a) On the top, is the longitudinal profile of one of the hybrid events selected with two of its matching simulated showers, for proton and iron. On the bottom is the ground signal for the same data event and simulations. Right: (b) Results for $R_\text{E}$ and $R_\mu$ for QGSJetII-04 and EPOS-LHC, for mixed and pure proton composition. The ellipses and the gray boxes are the statistic and systematic uncertainties, respectively.  }%
  \label{fig: muons2}
\end{center}
\end{figure}

At the Pierre Auger Observatory, we have collected a large number of hybrid events, for which we measured the longitudinal profile (LP) with the FD and the ground signal on the SD. For these events it is possible to simulate air showers with matching LP, in order to reduce the ground signal fluctuations event-by-event, and compare the simulated ground signal with the data. \\
In this study the data collected between 01 January and 31 December 2012 were also used. The hybrid events were selected using the cuts defined for the energy calibration of the SD \cite{Energy} and only events with $10^{18.8}<\hat{E}<10^{19.02}$ eV were used. The showers were simulated using SENECA\cite{Seneca}, with FLUKA\cite{Fluka} and for QGSJetII-04 or EPOS-LHC. For each hybrid data event, simulations with same geometry and energy were performed until 12 showers matched $X_{max}$ within one sigma of the real event. From the 12, only the 3 simulated showers that best matched the real LP were considered for full simulation and posterior data comparison. This was performed for proton and iron primaries and for QGSJetII-04 and EPOS-LHC. In figure \ref{fig: muons2}(a) top, one LP for a data event is plotted with a corresponding simulated LP for proton and iron. On the bottom is the ground signal for the same data event and simulations. It is clear that the real event has a higher signal than the two simulated events chosen to represent its longitudinal profile.\\
 The signal in the SD at $1000$ m, $S(1000)$, has different fractions coming from muons at different zenith angles. Knowing the evolution the EM and muonic fraction in $S(1000)$ with the angle, from simulation (see \cite{ICRC13-Farrar}), we can obtain a global rescaling factor $R_\text{E}$ and $R_\mu$ for the EM and muonic components respectively, where $S(1000)$ is then given by equation \ref{eq: S1000}. Since $S(1000)$ is a reconstructed parameter of each event, the EM and muon components are rescaled before recalculating the detector response for each simulated shower.
\begin{align}
S_{data}(1000) = S_{resc}(R_\text{E},R_\mu)_{i,j} \equiv R_\text{E} S_{\text{EM},i,j} + R^{\alpha}_\text{E} R_\mu S_{\mu,i,j}
\label{eq: S1000}
\end{align}
where $\alpha$ is the energy scaling of the muonic signal ($\approx0.89$ for both models), $i$ runs over each event in data and $j$ labels the primary.\\
In figure \ref{fig: muons2}(b), the results for $R_\text{E}$ and $R_\mu$ are plotted for QGSJetII-04 and EPOS-LHC, for mixed and pure proton composition. The mixed composition is the mix of $p$, He, N and Fe which best-fits the observed $X_{max}$ distribution.

The results obtained are written in table \ref{tab:results}, $R_\text{E}$ are very close to one, while $R_\mu$ shows a rescaling of about $30\%$ to $59\%$, exposing a deficit of muons in the models.
\begin{table}
\tbl{$R_\text{E}$ and $R_\mu$ results $\pm$stat. $\pm$syst. uncertainties.}
{\begin{tabular}{@{}cccc@{}}
\toprule
H. Model & $R_E$ & $R_\mu$ \tabularnewline
\colrule
QGSJetII-04 $p$ & $1.09\pm0.08\pm0.09$ & $1.59\pm0.17\pm0.09$ \\
QGSJetII-04 mixed & $1.00\pm0.08\pm0.11$ & $1.59\pm0.18\pm0.11$ \\
EPOS-LHC $p$ & $1.04\pm0.08\pm0.08$ & $1.45\pm0.16\pm0.08$\\
EPOS-LHC mixed & $1.01\pm0.07\pm0.08$ & $1.30\pm0.13\pm0.09$\tabularnewline
\botrule
\end{tabular}
}
\label{tab:results}
\end{table}

\section{Discussion}
In the muon analysis at the Pierre Auger Observatory, a muon deficit has been observed in the leading hadronic interaction models with respect to data, consistent for different methods.
The methods in section \ref{smooth} and \ref{multivariate} result in a muon signal in data around $1.3$ higher than QGSJetII-04 for $10^{19}$ eV and $\theta<60^{\circ}$. These values are compatible with iron primaries in both models, QGSJetII-04 and EPOS-LHC. For inclined showers (with $\theta>62^{\circ}$), in section \ref{Inclined}, data has around $1.8$ more muons than QGSJetII-03 at $10^{19}$ eV. This corresponds to about $1.5$ in relation to QGSJetII-04. The inclined results are marginally comparable to the prediction for iron showers, but the relative number of muons ($R_\mu$) seems to increase with energy, while for the models it decrease. These discrepancies with models could be related to an incorrect energy within the $22\%$ systematic uncertainty of the energy scale or problems in the simulation of the hadronic and muonic shower components.\\
On the hybrid studies, the muon rescaling in $10$ EeV air showers ($E_{\text{CM}}=137$ TeV)  is a factor 1.3 to 1.6 larger than predicted using the leading hadronic interaction models tuned to fit the LHC data and lower energy accelerator data. 
However the possibility of having a heavy composition, similar or heavier than iron, is in clear contradiction with the measurements of the depth of shower maximum \cite{Xmax}.

\section*{Acknowledgments}
Thanks to LIP and FCT for the PhD grant and for giving me the chance to participate in a top research team. Also, this would not have been possible without the strong commitment and effort of the technical and administrative staff in 
Malarg\"ue.  

\bibliographystyle{ws-procs975x65}

\end{document}